\documentclass[sigconf,screen]{acmart}

\usepackage[ruled,vlined,linesnumbered]{algorithm2e}
\usepackage{algpseudocode}
\usepackage{color}
\usepackage{multirow}
\usepackage{amsfonts}
\usepackage{xspace}
\usepackage{colortbl}
\usepackage{makecell}
\usepackage{microtype}

\newcommand{\tool}{BehAVExplor\xspace}
\newcommand{\miner}{BehaviorMiner\xspace}

\usepackage{enumitem}

\AtBeginDocument{%
  }

\setcopyright{acmlicensed}
\acmPrice{15.00}
\acmDOI{10.1145/3597926.3598072}
\acmYear{2023}
\copyrightyear{2023}
\acmSubmissionID{issta23main-p126-p}
\acmISBN{979-8-4007-0221-1/23/07}
\acmConference[ISSTA '23]{Proceedings of the 32nd ACM SIGSOFT International Symposium on Software Testing and Analysis}{July 17--21, 2023}{Seattle, WA, USA}
\acmBooktitle{Proceedings of the 32nd ACM SIGSOFT International Symposium on Software Testing and Analysis (ISSTA '23), July 17--21, 2023, Seattle, WA, USA}
\received{2023-02-16}
\received[accepted]{2023-05-03}

\begin{document}
\SetEndCharOfAlgoLine{}

\title{BehAVExplor: Behavior Diversity Guided Testing for Autonomous Driving Systems}

\author{Mingfei Cheng}
\email{mfcheng.2022@phdcs.smu.edu.sg}
\affiliation{
  \institution{Singapore Management University}
  \country{Singapore}
}

\author{Yuan Zhou}\authornote{Corresponding author}
\email{y.zhou@ntu.edu.sg}
\affiliation{
  \institution{Nanyang Technological University}
  \country{Singapore}
}

\author{Xiaofei Xie}
\email{xfxie@smu.edu.sg}
\affiliation{
  \institution{Singapore Management University}
  \country{Singapore}
}
\renewcommand{\shortauthors}{Mingfei Cheng, Yuan Zhou, and Xiaofei Xie}

\begin{abstract}
Testing Autonomous Driving Systems (ADSs) is a critical task for ensuring the reliability and safety of autonomous vehicles. Existing methods mainly focus on searching for safety violations while the diversity of the generated test cases is ignored, which may generate many redundant test cases and failures. Such redundant failures can reduce testing performance and increase failure analysis costs. 
In this paper, we present a novel behavior-guided fuzzing technique (\tool) to explore the different behaviors of the ego vehicle  (i.e., the vehicle controlled by the ADS under test) and detect diverse violations. 
Specifically, we design an efficient unsupervised model, called \miner, to characterize the behavior of the ego vehicle. 
\miner extracts the temporal features from the given scenarios and performs a clustering-based abstraction to group behaviors with similar features into abstract states. 
A new test case will be added to the seed corpus if it triggers new  behaviors (e.g., cover new abstract states). Due to the potential conflict between the behavior diversity and the general violation feedback, we further propose an energy mechanism to guide the seed selection and the mutation. The energy of a seed quantifies how good it is. We evaluated \tool on Apollo, an industrial-level ADS, and LGSVL simulation environment.  Empirical evaluation results show that \tool can effectively find more diverse violations than the state-of-the-art.
\end{abstract}




\maketitle

\section{Introduction}
Autonomous Vehicles (AVs) play an important role in smart cities, which can relieve traffic congestion and reduce accidents in intelligent transportation systems.
In AVs, human drivers are replaced by Autonomous Driving Systems (ADSs), which perform various driving tasks, such as perception, localization, planning, and control.
However, ADSs are vulnerable to various issues such as wrong implementation of algorithms, incorrect condition logic, and unsuitable configurations~\cite{garcia2020comprehensive}.
Hence, before an AV is deployed to the real world, its ADS should be tested sufficiently to guarantee safety under possible scenarios in the real world.

A scenario describes the relevant characteristics, activities, and$/$or goals of the ego vehicle (i.e., the AV controlled by the ADS under test) and other objects (e.g., pedestrians and other vehicles).
Scenarios can be designed in different categories, from the most abstract to the least: functional scenarios, which describe roads and objects in linguistic notation; logical scenarios, which define the parameters and their ranges in scenarios; and concrete scenarios, where concrete values of the parameters are given \cite{menzel2018scenarios}.
Test cases of ADS are executable concrete scenarios together with oracles that the behavior of the ADS must satisfy.
ADS testing aims to discover critical scenarios, i.e., concrete scenarios violating the oracles \cite{neurohr2020fundamental}.

The methods of ADS testing can be generally divided into on-road testing and simulation-based testing.
Even though on-road testing is necessary, it is usually costly and, most importantly, difficult to cover enough situations that AVs should be able to safely react to (e.g., different accident scenarios).
In contrast, leveraging the high-fidelity simulators, such as LGSVL~\cite{rong2020lgsvl} and CARLA~\cite{dosovitskiy2017carla}, simulation-based testing allows developers and engineers to systematically assess ADSs against a broader range of concrete scenarios that may occur on the road. 
However, due to the open and unpredictable road conditions (e.g., weather, traffic flow, vehicle behaviors), there are infinite concrete scenarios. Hence, a key challenge of AV testing is how to effectively discover diverse scenarios.

\begin{figure}[!t]
    \centering
    \includegraphics[width=1.0\linewidth]{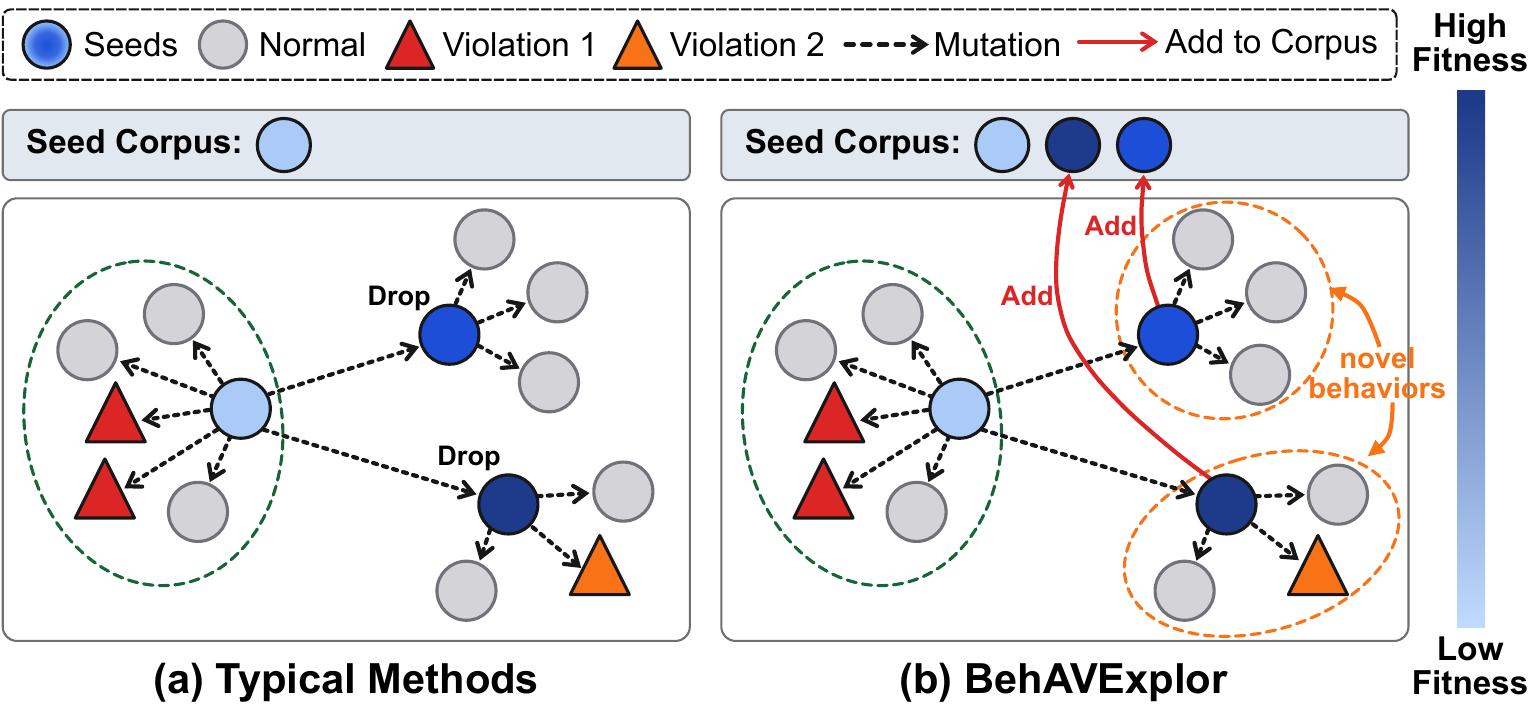}
    \caption{Illustration of (a) existing ADS testing methods and (b) our \tool. The lengths of dash lines represent the behavior similarity.}
    \label{fig:intro_idea}
\end{figure}

Several ADS testing methods have been proposed to generate critical scenarios efficiently, such as data-driven methods~\cite{Bashetty20DeepCrashTest,ding2020cmts,paardekooper2019automatic}, surrogate models~\cite{beglerovic2017testing,mullins2018accelerated}, and guided searching methods~\cite{av_fuzzer,icse_samota,tse_adfuzz,gambi2019automatically}. 
They usually design risk evaluation metrics (e.g., time to collision and proportion of stopping distance)~\cite{mahmud2017application} to guide the generation of critical scenarios. However, the existing techniques mainly focus on discovering failed scenarios (i.e., violations), while the diversity of generated scenarios is not well considered. As shown in Figure~\ref{fig:intro_idea}(a), 
some seeds that have the potential to generate new critical scenarios (leading to new violated behaviors) are ignored without considering the behavior diversity of the ego vehicle.
This can cause the generation of redundant critical scenarios, which limits the effectiveness of testing and increases the cost of bug analysis.
Especially, the simulator usually takes a long time to run generated scenarios (ranging from seconds to tens of seconds). Hence, it is important to generate scenarios with diverse ego behaviors, which can test the target ADS better and expose various issues.

In this paper, we propose a novel fuzzing approach that 
aims to improve the diversity of the generated scenarios by incorporating the guidance of the behavior diversity
(e.g., orange circles in Figure~\ref{fig:intro_idea}(b) are added into the seed corpus). There are some challenges in generating critical scenarios that have diverse behaviors:
1) Considering the complex traffic conditions and the temporal motion of vehicles, it is challenging to characterize the behavior diversity of given scenarios. 2) The two objectives (i.e.,  behavior diversity and violation) may conflict. Focusing more on diversity may reduce the likelihood of detecting violations. Conversely, focusing more on violations may lack behavior diversity. How to balance the two objectives becomes another challenge.

Facing the above challenges, in this paper, we propose a diversity and  violation-guided fuzzing framework, \tool, to generate critical scenarios in which the ego vehicle has diverse behaviors, i.e., diverse violations. 
Specifically, to address the first challenge, we propose an abstraction method based on clustering that characterizes similar behaviors (from a number of scenarios) into abstract states. Given a set of test cases, \tool first collects driving traces of the ADS, where each trace contains a sequence of concrete states. Since the states in a trace may be sparse, linear interpolation is performed to add states in the trace. To capture the temporal behavior of the ego vehicle, \tool then adopts the sliding window on each trace to extract the temporal features from consecutive states. Finally, \tool clusters the temporal features as abstract states that represent different behaviors of the ego vehicle. For a given scenario, \tool extracts an abstract trace consisting of abstract states. We calculate the distances between the abstract trace of a new test case and those of the existing test cases to measure its diversity. Moreover, to evaluate the violation potential of the scenario, a criticality measurement called violation degree is defined, which evaluates how far the scenario is from violating the given specifications. 
To address the second challenge, we propose an energy mechanism that maintains energy for the seeds in the seed corpus. The energy of a seed measures the potential of the seed, which is calculated based on its failure triggering rate, violation degree, and selection frequency. We further propose an energy-based seed selection strategy and an adaptive mutation that can improve the testing performance in terms of the two objectives.

We evaluated the effectiveness of \tool on the LGSVL simulator and Baidu Apollo ADS. Specifically, we compared \tool with three baselines, i.e., random testing, AVFuzzer~\cite{av_fuzzer} and SAMOTA~\cite{icse_samota}. The evaluation results show that \tool discovers more violations (an increase of 61.8\% on average) than the baselines (the best one is AVFuzzer). The violations discovered by \tool are more diverse, e.g., an average of 7.2 more unique violations are discovered than the best of the baselines. We further demonstrated the usefulness of the diversity feedback and the energy mechanism.

In summary, this paper makes the following contributions:
\begin{enumerate}[leftmargin=*, itemsep=0pt, topsep=0pt, parsep=0pt]
\item We propose a behavior-guided fuzzing method, \tool, to discover critical scenarios with diverse behaviors. 
\item We propose a clustering-based abstraction to  characterize the ego behavior in a scenario.
\item We propose an energy-based seed selection strategy and an adaptive mutation to improve the testing performance.
\item We evaluate \tool on Baidu Apollo, one of the state-of-the-art open-source and production-scale ADSs, and discover 16 unique violations. All detailed results as well as the source code can be found in~\cite{ourweb}.
\end{enumerate}

\section{Background and Problem Definition}
\subsection{Autonomous Driving Systems}
Autonomous Driving Systems (ADSs) are the brain of Autonomous Vehicles (AVs). Existing ADSs can mainly be divided into two categories: End-to-End (E2E) driving models and multi-module ADSs~\cite{tampuu2020survey}. E2E driving models process sensor data to generate control decisions by using a single deep learning model. Although recent advances in Deep Neural Networks (DNNs) have led to the development of E2E driving models, such as PilotNet~\cite{bojarski2016end,bojarski2017explaining} and OpenPilot~\cite{openpilot}, E2E driving models, which require a large amount of training data, still perform poorly in generalization and industrial performance. In contrast, multi-module ADSs are preferred in the industry for their reliable performance in handling various scenarios, such as Baidu Apollo~\cite{apollo} and Autoware~\cite{autoware}. Therefore, in this paper, we choose to test multi-module ADSs. 

% \begin{figure}[t!]
%     \centering
%     %\setlength{\abovecaptionskip}{-0.05cm}
%     \includegraphics[width=0.9\linewidth]{figs/BG-ADS.pdf}
%     \caption{Overview of the typical multi-module ADS.}
%     \label{fig:bg_ads}
%     %\vspace{-0.9cm}
% \end{figure}

A typical multi-module ADS, such as Baidu Apollo \cite{apollo}, usually consists of localization, perception, prediction, planning, and control modules to process rich sensor data and make decisions (the typical architecture is given in \cite{ourweb}). The localization module provides the location of the AV by fusing multiple input data from GPS, IMU, and LiDAR sensors. The perception module takes camera images, LiDAR point clouds and Radar signals as inputs to detect the surrounding environment (e.g., traffic lights) and objects (e.g. other vehicles and pedestrians) by mainly using deep neural networks. The prediction module is responsible for tracking and predicting the trajectories of all surrounding objects detected by the perception module. Given the results of perception and prediction modules, the planning module then generates a local collision-free trajectory for the AV. Finally, the control module converts the planned trajectory to vehicle control commands (e.g., steering, throttle, and braking) and sends them to the chassis of the AV. In this paper, we conduct ADS testing on a simulation platform, where the ADS connects to a simulator via a communication bridge, receives sensor data from the simulator, and sends the control commands to the AV in the simulator. 

\subsection{Scenario Description}
\textbf{Scenario Definition.} 
A scenario is a temporal sequence of scenes, where each scene is a snapshot of the environment including the scenery and dynamic objects~\cite{scenario_definition}.
Hence, a scenario can be described by a series of configurable static and dynamic attributes. 
Static attributes set the static objects of the scenario, such as the  map, weather, time of the day, and so on.
Dynamic attributes define the traffic flow in a scenario, containing the set of states (e.g., position and orientation), trajectories, and behaviors (e.g., speed) of dynamic objects, e.g., Non-Player Character (NPC) vehicles (i.e., other vehicles except the ego one). 
Usually, a complex scenario can consist of hundreds and even thousands of attributes extracted from the Operational Design Domains (OODs)~\cite{thorn2018framework}.
It is impossible to describe and evaluate all attributes. 
Facing different testing purposes, researchers focus on some specific attributes. 
For example, to evaluate the robustness of the perception module under different weather conditions, researchers may focus more on the configurations of different weathers, such as rain and fog~\cite{wang2022performance,sun2019convolutional,cheng2018sparse}.
To evaluate the safety of an ADS, researchers focus more on the configurations of the trajectories and behaviors of the NPC vehicles~\cite{av_fuzzer,icse_samota,tang2021systematic}.

Motivated by these preliminary studies, considering ADS safety testing, we also focus on the configurable attributes for each NPC vehicle, i.e., the route described by the initial and target positions and the trajectory described by a set of waypoints, to search for diverse and critical scenarios. 
Different from the existing work~\cite{av_fuzzer, icse_samota} which only mutates the waypoints of each NPC vehicle without specifying the route explicitly, we mutate both the NPC's route and its corresponding waypoints, each of which is formulated by the \textit{position} (denoted as $p$) and \textit{velocity} (denoted as $v$) of the vehicle. In this way, we aim to search for critical scenarios more efficiently. 

\noindent \textbf{Scenario Observation.} Scenario observation \cite{scenario_definition} is a sequence of scenes, which are frames including the states of the objects in a scenario.
After executing a scenario $s$ with a running duration of $t_{sim}$, the corresponding trace of the scenario observation $\textbf{X}_s =\{ \textbf{x}_{s}^{t}|t=1,2,...,t_{sim}\}$ can be obtained from the simulator and the ADS. Each frame $\textbf{x}_{s}^{t}=\{ [time^t, (p_{k}^{t},h_{k}^{t},v_{k}^{t},a_{k}^{t})]|k=0,1,...,n_{npc}\}$ contains the timestamp of the frame, the center position $p_{k}^{t}$, heading $h_{k}^{t}$, velocity $v_{k}^{t}$ and the acceleration $a_{k}^{t}$ of the vehicle $k$, where  $k=0$ represents the ego vehicle.

\subsection{Specifications for ADSs \label{sec_ads_violation}}
ADSs should satisfy various specifications. 
In this paper, we choose three essential safety specifications~\cite{safety_first,icse_samota}: collision avoidance, not hitting illegal lines, and reaching the destination. 
ADS testing target is to explore scenarios violating the three specifications, which results in different kinds of violations.
The formulae of the three specifications are given as follows. 

\noindent \textbf{R1: Collision Avoidance.} Given a scenario $s$ and its execution observation trace $\textbf{X}_{s}=\{\textbf{x}_{s}^{t}|t=1,2,...,t_{sim}\}$, let $D_b(p_{0}, s)$ denote the minimum distance between the ego vehicle and the NPC vehicles in $s$ during the execution, i.e.,
\begin{equation}
    D_b(p_{0}, s) = \min(\{D_{b2b}(p_{0}^{t}, p_{k}^{t})|1 \leq k \leq n_{npc}, 1 \leq t \leq t_{sim}\})
\end{equation}
where $D_{b2b}(p_{0}^{t}, p_{k}^{t})$ computes the shortest distance between the bounding boxes of the ego vehicle and vehicle $k$ at the time instant $t$, whose positions are $p_{0}^{t}$ and $p_{k}^{t}$, respectively.
Hence, the violation of collision avoidance can be described as:
\begin{equation}\label{eqn:collision}
    D_b(p_{0}, s) = 0
\end{equation}

\noindent \textbf{R2: Not Hitting Illegal Lines.} 
In this paper, we consider a critical traffic rule: the ego vehicle should not hit illegal lines (e.g., double yellow lines and curb lines). 
Given a simulation observation trace $\textbf{X}_{s}$, the minimum distance between the ego vehicle and the illegal lines during the execution is defined as:
\begin{equation}
    D_l(p_0, s) = \min\{D_{b2l}(p_{0}^{t}, l)|1 \leq t \leq t_{sim}, l \in \mathbb{L} \}
\end{equation}
where $D_{b2l}(p_{0}^{t}, l)$ calculates the shortest distance between the bounding box of the ego car and the illegal line $l$, and  $\mathbb{L}$ is the set of illegal lines in the testing region.
Hence, its violation can be defined as:
\begin{equation}\label{eqn:rule}
    D_l(p_0, s) = 0
\end{equation}

\noindent \textbf{R3: Reaching Destination.} Following existing works \cite{icse_samota, css_drivefuzzer}, reaching the assigned destination is one of the most important requirements for an ADS. 
The violation of reaching the destination is not expected and should be forbidden.
We use the distance between the last position of the ego vehicle, i.e., $p_{0}^{t_{sim}}$, and the destination $p_{dest}$ in $\textbf{X}_s$ to determine whether the scenario is passed. The detail condition is defined by:
\begin{equation}\label{eqn:destination}
    D_{p2p}(p_{0}^{t_{sim}}, p_{dest}) \leq threshold.
\end{equation}
Because there will be some estimation errors in the simulator (e.g. length of agents), we consider a scenario to be passed if the distance is within a certain threshold. The threshold is set to 1 in our paper.

\section{Approach}

% In this section, we first give an overview of \tool in Section~\ref{sec_method:overview}, followed by the design of a behavior mining model \miner (Section \ref{sec_method:behavior_miner}) and an adaptive strategy for seed selection and mutation (Section \ref{sec_method:energy_mechanism}). 
\begin{figure*}[ht!]
    \centering
	\includegraphics[width=0.8\textwidth]{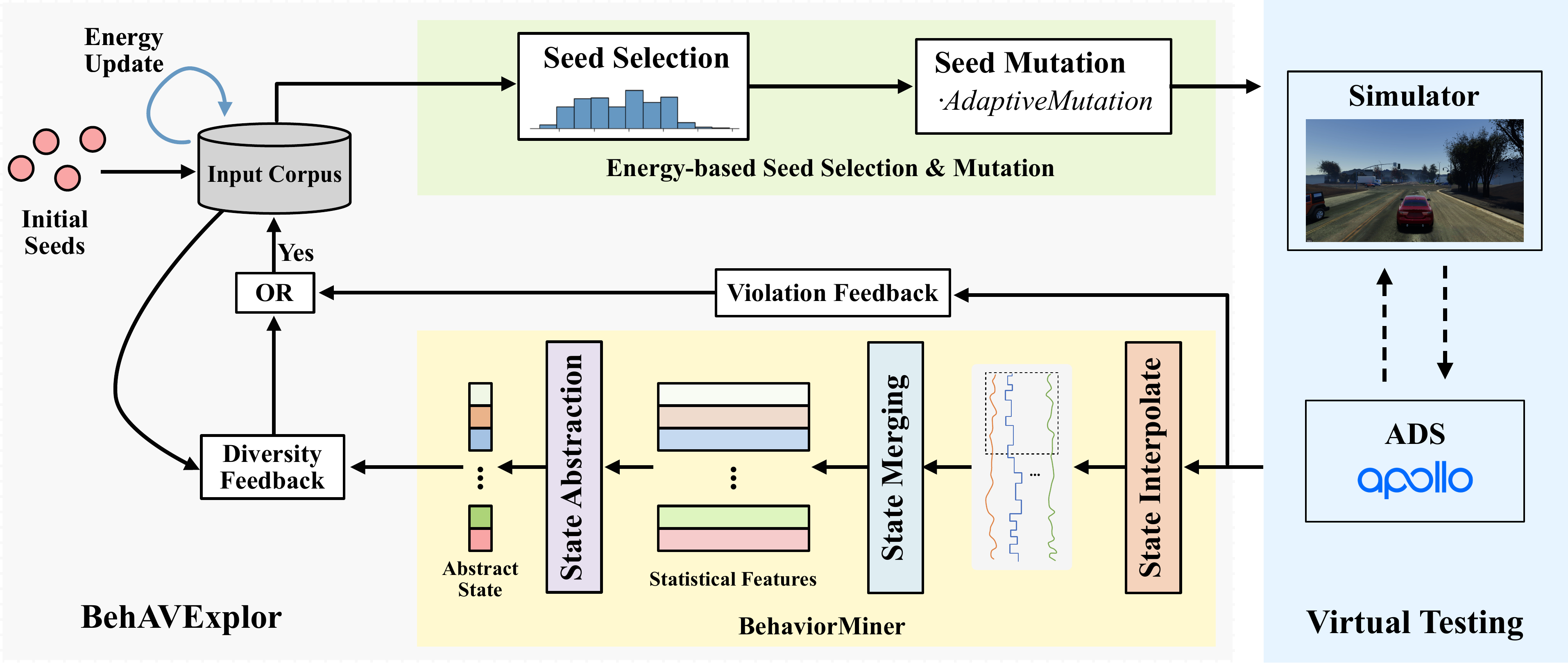}
	\caption{Overview of \tool.}
 	\label{fig: bediv_overview}
\end{figure*}
\subsection{Overview \label{sec_method:overview}}
Figure \ref{fig: bediv_overview} illustrates the overview of \tool. Like a general fuzzer, \tool maintains a seed input corpus that stores ``interesting'' seed inputs that are helpful in identifying failed or diverse test cases. At each iteration, \tool adopts an \emph{energy} mechanism to select a seed with higher energy from the input corpus. The {\emph{energy}} of a seed quantifies how well it can generate failed test cases. Then an adaptive mutation strategy is proposed to generate new test cases by mutating the selected seed. 

To search for diverse and critical test cases, \tool defines the behavior diversity and scenario criticality as the fuzzing feedback to select ``interesting'' test cases that can increase the diversity or violation degree. Specifically, the new mutants will be fed into the target ADS to generate observation traces. \tool characterizes the behavior of each mutant through \miner. The diversity is measured based on the difference between the behavior of the new input and the behaviors of existing seeds in the seed corpus. Moreover, the general violation functions (e.g., collision, hitting illegal lines) regarding AV failures are defined to evaluate the criticality of the mutant. The mutants with new behaviors or better violation degrees will be added to the input corpus.

Algorithm \ref{algo:BeDivFuzzer} shows the main algorithm of \tool. 
It takes a set of initial seeds \textbf{Q} and a target ADS \textit{P} as inputs and outputs benign scenarios  $\textbf{Q}$ and failed tests $\textbf{F}$. There are also some configurable parameters, including the number of clusters $K$ in \miner, the diversity threshold $\theta_{c}$, and the mutation threshold $\theta_{e}$.

\begin{algorithm}[!t]
\small
    \SetKwInOut{Input}{Input}
    \SetKwInOut{Output}{Output}
    \SetKwInOut{Para}{Parameters}
    \SetKwComment{Comment}{/*}{*/}
    \SetKwFunction{AdaptiveMutation}{{\textbf{AdaptiveMutation}}}
    \SetKwFunction{Select}{{\textbf{Select$_E$}}}
    \SetKwFunction{BehaviorMiner}{{\textbf{BehaviorMiner}}}
    \SetKwFunction{EnergyInitializer}{{\textbf{EnergyInitializer}}}
    \SetKwFunction{EnergyUpdater}{{\textbf{EnergyUpdater}}}
    \SetKwFunction{Simulator}{\textbf{Simulator}}
    
    \Input{Initial seed corpus $\textbf{Q}$, Target ADS $P$
    }
    \Output{Diverse test cases $\textbf{Q}$, 
            Failed tests $\textbf{F}$ }
    \Para{Cluster number in \miner $K$, 
           Diversity threshold $\theta_{c}$,
           Adaptive mutation threshold $\theta_{e}$ 
           }
           
     $\textbf{F} \gets \{\}$, $\textbf{Y} \gets \{\}$ 
    
    \For{$s \in Q$}{$E_s\gets 1$\Comment*[r]{Initialize energy for each seed}} 
    \BehaviorMiner.update($K$, \textbf{Q}) \Comment*[r]{Initialize BehaviorMiner}
    
    \Repeat{given time budget expires}
    {
        $s \gets \Select(Q)$ \Comment*[r]{Energy-based seed selection} 
        $s' \gets \AdaptiveMutation(s, E_s, \theta_{e})$ \Comment*[c]{adaptive mutation} 
        $\textbf{X}_{s'} \gets \Simulator(s', P)$ \Comment*[c]{execute the mutant} 
        $\hat{\textbf{H}}_{s'}, O_{{s'}}, R_{{s'}} \gets$ \BehaviorMiner.analyze($\textbf{X}_{s'}$) \\
        \eIf{$R_{s'}$ = FAILURE}{
            $\textbf{F} \gets \textbf{F} \cup \{s'\}$ 
        }{
            \If{$distance(\hat{\textbf{H}}_{s'}, \textbf{Y}) > \theta_{c}$ or $O_{s'} < O_{s}$}{
                $\textbf{Q} \gets \textbf{Q} \cup \{s'\}$,\quad
                $\textbf{Y} \gets \textbf{Y} \cup \{\hat{\textbf{H}}_{s'}\}$ \\
                $E_{{s'}}  \gets \EnergyInitializer(O_{s}, O_{{s'}}, R_{{s'}})$  
            }

            \If{$distance(\hat{\textbf{H}}_{s'}, \textbf{Y}) > \theta_{c}$}{
                \BehaviorMiner.update($K$, $\textbf{Q}$) \Comment*[l]{update clusters}
            }
        }
        $E_{{s}} \gets \EnergyUpdater(O_{{s'}}, O_{{s}}, R_{s'})$
    }
    
    \Return $\textbf{Q}, \textbf{F}$

\caption{BehAVExplor Algorithm}
\label{algo:BeDivFuzzer}
\end{algorithm}

\textbf{Initialization}. \tool begins with initializing an empty failed corpus $\textbf{F}$ and an empty behavior set $\textbf{Y}$ (i.e., the set of abstract traces) (line 1). 
Each seed in the corpus is initialized with the same energy (lines 2-3).
Then \miner initializes its cluster model based on the initial seeds $\textbf{Q}$ (line 4, detailed at Section \ref{sec_method:behavior_miner}).

\textbf{Fuzzing.} During the fuzzing process, \tool repeats feedback (including behavior diversity and specification violation degree) guided fuzzing until the stopping criterion is satisfied (lines 5-19). 
In each iteration, \tool first adopts an energy-based mechanism to select a seed $s$ from the seed corpus $\textbf{Q}$ (line 6) and adaptively mutate it such that a new test case $s'$ is generated (line 7).
Then, the new mutant $s'$ is executed in the simulator with the target ADS $P$ and collects its observation trace $\textbf{X}_{s'}$  (line 8).
According to $\textbf{X}_{s'}$, \miner extracts its abstract behavior trace $\hat{\textbf{H}}_{s'}$, violation degree $O_{s'}$, and execution result $R_{s'}$ (line 9, detailed in Section \ref{sec_method:behavior_miner}). The mutant is added to the failed test set if its result is a failure (lines 10-11). Otherwise, \tool decides whether to retain the mutant $s'$ (Lines 13-14) and calculates the energy of the new seed $s'$ according to $R_{s'}$, $\hat{\textbf{H}}_{s'}$ and $O_{s'}$ (Line 15).
Specifically, $s'$ is kept if (1) it triggers new behaviors, i.e., the distance between $\hat{\textbf{H}}_{s'}$ and the existing behaviors $\textbf{Y}$ is larger than the diversity threshold $\theta_{c}$, or (2) the violation degree of $s'$ is less than that of its parent $s$ (i.e., $O_{s'} < O_{s}$).
Moreover, if new behaviors are discovered, \miner updates its behavior cluster model based on the new seed corpus $\textbf{Q}$ (lines 16 and 17).
Finally, the energy of $s$ is updated according to the analysis of $s'$ (line 18).
Note that the violation degree represents the fitness value with regard to a failure specification such as collision or hitting illegal lines. The two kinds of feedback work together such that \tool can find more diverse failures.
The algorithm ends by returning $\textbf{Q}$ and $\textbf{F}$ (line 20). 

\subsection{Diversity and Violation Feedback\label{sec_method:behavior_miner}}
In general, \tool aims to discover scenarios that can cause \textit{violations} and have \textit{diverse} behaviors.  For example, the ego vehicle may crash with different behaviors such as overtaking, accelerating, or turning the corner.  While existing techniques mainly focus on how to detect violations effectively, they do not fully consider the diversity of ego behaviors. 
As a result, it may generate redundant violations with low diversity.

\begin{figure}[!t]
    \centering
    \includegraphics[width=\linewidth]{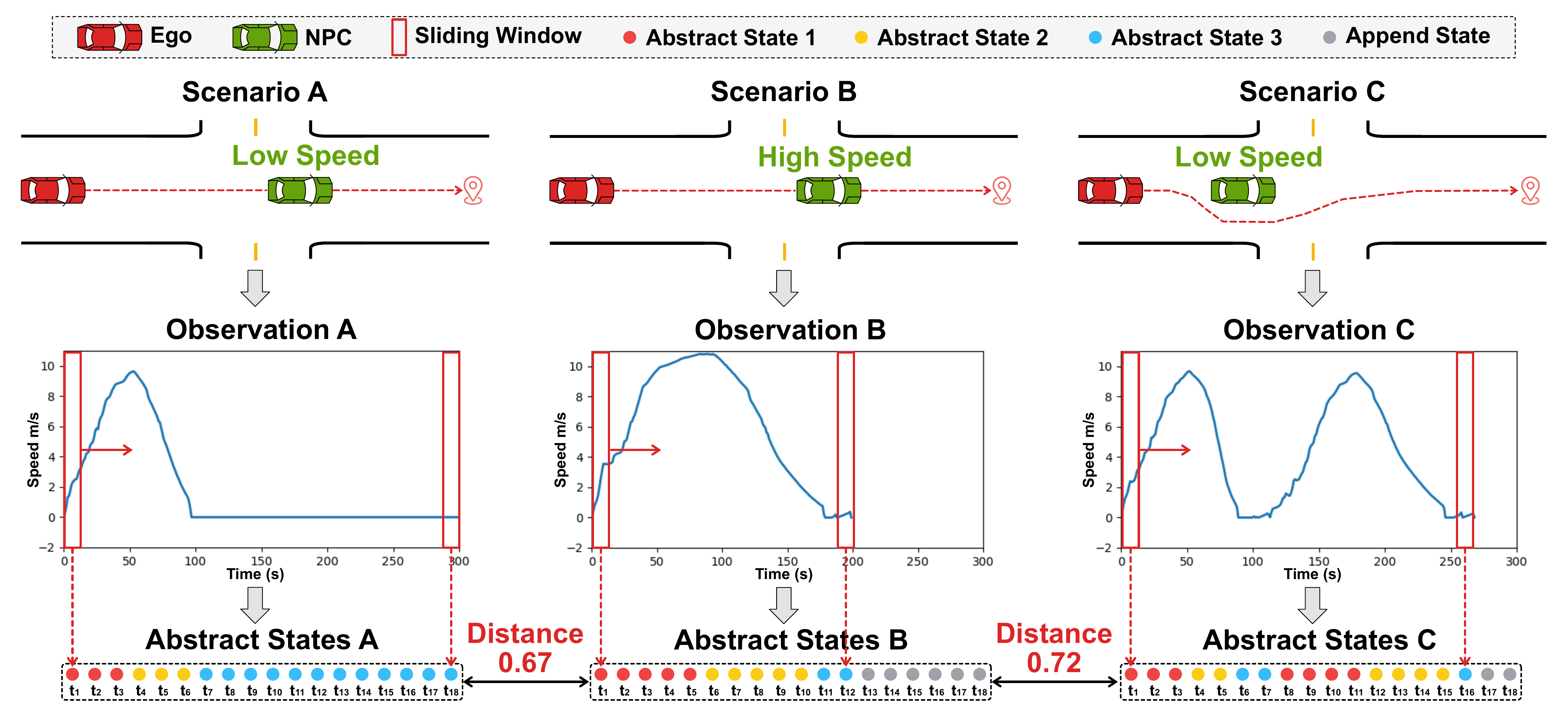}
    \caption{Examples of behavior-based diversity representation. Scenario A: ego follows a slow NPC. Scenario B: a normal scenario. Scenario C: ego bypasses a slow NPC. }
    \label{fig:method_div}
    %\vspace{-0.0cm}
\end{figure}

\tool incorporates feedback from diversity and violation degree. A key challenge is how to measure the diversity of ego behaviors. To this end, we propose a clustering-based behavior mining method, \miner, to characterize ego behaviors. Then, a distance-based metric is designed to calculate the similarity of ego behaviors in two scenarios. 
Figure \ref{fig:method_div} shows three scenarios to illustrate the idea of our behavior guidance. The observation of each scenario (e.g., speed) will be processed by \miner to obtain a sequence of abstract states. As shown at the bottom part of Figure \ref{fig:method_div}, the difference in the three abstract state sequences is used to measure the behavior difference (i.e., the diversity). 
To evaluate how far the existing scenario is from violating a given specification, we adopt some commonly used metrics for \textit{collision}, \textit{hitting illegal liens}, and \textit{reaching destination} (see Section~\ref{sec_ads_violation}).

% %\vspace{-0.23cm}
% %\vspace{-\textheight}
\subsubsection{Behavior Diversity}
Given a test case $s$ that is fed into the target ADS, the returned observation trace is denoted as $\textbf{X} = \{\textbf{x}^{1}, \textbf{x}^{2}, ..., \textbf{x}^{N}\} \in \mathbb{R}^{M \times N}$, where $\textbf{x}^{n} = (x_{1}^{n}, x_{2}^{n}, ..., x_{M}^{n})$ represents the state in the $n$-th frame of $\textbf{X}$ containing some important motion attributes of the ADS (e.g., timestamp, velocity, heading, etc.), $x_{i}^{n}$ means the value of the attribute $i$ in the $n$-th frame, $M$ is the number of attributes, and $N$ denotes the number of frames in the trace $\textbf{X}$. 
%(i.e., a scenario)

Given two scenarios, a simple idea is to measure the behavior difference of their observation traces. However, due to the high dimension of each frame, the large differences among the values of attributes, and the different lengths of two traces, it is difficult to precisely capture the diversity on the raw traces. To address this problem, we propose an abstraction-based method \miner that clusters similar behaviors (e.g., states in frames) into an abstract state. Finally, we measure the behavior diversity based on the abstract trace which is a sequence of abstract states.

\miner performs the abstraction based on a given set of test cases (i.e., scenarios in the seed corpus). The abstraction could characterize the  behaviors of existing test cases.  Specifically, \miner adopts \textit{state interpolation} and \textit{state merging} to extract temporal features of ego behaviors from each observation trace, which are used for the subsequent \textit{state abstraction}. 

\noindent \textbf{State Interpolation.} 
Sate interpolation is to address the problem when frames in a trace are not uniformly sampled from the simulator, which may result in extracting unstable features. For example, the time intervals between two consecutive frames in an \textit{asynchronous} sampling may not be the same, making the behavior measurement not accurate. 
Hence, \tool performs linear interpolation to generate an approximate trace with an equal time step.
In detail, \tool first approximates a piecewise linear function (i.e., using a linear function to approximate the trace between two consecutive states) for $\textbf{X}$ and then samples a set of states with an equal time step. 
Finally, a new trace $\overline{\textbf{X}} \in \mathbb{R}^{M \times N'}$ is generated, where $N'$ is the number of states in the new trace, and the time intervals between any consecutive states are the same. 

\noindent \textbf{State Merging.}  
Intuitively, the ego behaviors are represented by a sequence of frames such as accelerating, turning the corner and overtaking. An individual state only captures the static information from a frame, which is not enough to capture the ego behaviors~\cite{behavioral_pattern, behavioral_pattern_segment}. 
Hence, given a trace $\overline{\textbf{X}}$, we use a sliding window $w$ to merge $w$ consecutive states as a state sequence. For example, for the state $\overline{\textbf{x}}_i$,  we extract $(\overline{\textbf{x}}_i, \ldots, \overline{\textbf{x}}_{i+w-1})$ that captures the temporal features. Note that, if the remaining states are less than $w$, we pad the sequence with the last state.

We define a function $f$ that merges a state sequence $(\overline{\textbf{x}}_i, \ldots,$ $ \overline{\textbf{x}}_{i+w-1})$ as a merged state $\textbf{h}_i$  by extracting statistical features:
\begin{equation}
\textbf{h}_i = f(\overline{\textbf{x}}_i, \ldots, \overline{\textbf{x}}_{i+w-1})
\end{equation}

The features in the merged state $\textbf{h}_i$ is a vector with the size $D\times M$ containing the values of different statistical measures, where $M$ is the number of attributes in each concrete state and $D$ is the number of statistical features we collected. The eight statistical measures are collected from each attribute including \textit{mean}, \textit{minimum}, \textit{maximum}, \textit{\textit{mean\_change}}, \textit{mean\_abs\_change}, \textit{variance}, \textit{non-linearity}, and  \textit{time series complexity} of each attribute sequence. The detailed explanation and calculation can be found on our website~\cite{ourweb}.

Finally, we extract a new trace $\textbf{H} =\{\textbf{h}_1,\ldots, \textbf{h}_{N'}\}\in \mathbb{R}^{T \times N'}$ containing a sequence of merged states from an observation trace~$\textbf{X}$.

\noindent \textbf{State Abstraction.} Based on the merged states, we adopt a clustering based abstraction $\sigma$ such that similar behaviors (represented by features in merged states) are grouped into the same abstract state:
\begin{equation}
\hat{\textbf{h}}=\sigma(\textbf{h})
\end{equation}
Specifically, we first extract a set of traces $(\textbf{H}_1, \textbf{H}_2, \ldots)$ from a given set of test cases (e.g., from seed corpus), where each trace contains a sequence of merged states. Then we adopt a clustering algorithm (K-Means~\cite{scikit-learn} used in this paper) to obtain a classifier $\sigma$ that clusters all merged states to $K$ abstract states. Note that each abstract state can be represented by a unique cluster ID. Finally, for a trace $\textbf{H}$, we obtain an abstract trace $\hat{\textbf{H}}\in \mathbb{Z}^{1 \times N'}$ by mapping the merged states to abstract states.

\noindent \textbf{Diversity Measurement.} 
To provide diversity feedback to the fuzzer, we need to measure whether a new mutant $s'$ can trigger new behaviors compared with the existing test cases. 
Distance is widely applied to measure diversity \cite{icse_guided, fse_deepstellar}. Specifically, we calculate a minimum distance as:
\begin{equation}
    d_{{s'}} = \min_{s\in \textbf{Q}} dis(\hat{\textbf{H}}_{s'}, \hat{\textbf{H}}_{s}).
\end{equation}
The larger the $d_{{s'}}$, the more diverse the ego behaviors in $s'$. We configure a pre-defined threshold $\theta_{c}$ to determine whether new behavior is discovered by the mutant $s_i$ (i.e., line 14 in Algorithm~\ref{algo:BeDivFuzzer}).

The execution time of different scenarios is an important factor when we measure behavior differences. For example, in Figure~\ref{fig:method_div}, the ego in Scenario A runs at a very slow speed and generates a longer trace, showing different behaviors with Scenarios B and C. 
Note that the grey states in Scenarios B and C are stop states, which means that the ego has arrived and stopped at the destinations. 
Therefore, we need to take the execution time into consideration when calculating the distance between two traces. 
So some existing time sequence distance metrics, such as Dynamic Time Warping (DTW) \cite{kdd_dtw}, may not work well in our situation. For example, the DTW distance between Scenario A and Scenario B in Figure \ref{fig:method_div} will be 0 even though their behaviors are different. 
In this paper, we design a distance metric based on the Hamming distance~\cite{hamming1986coding}, which computes the behavior distance between two sequences of abstract states in the same time scale. The distance is calculated by: 
\begin{equation}
    dis(\hat{\textbf{H}}_{s'}, \hat{\textbf{H}}_{s}) = \frac{Hamming(\hat{\textbf{H}}_{s'}, \hat{\textbf{H}}_{s}) + |len(\hat{\textbf{H}}_{s}) - len(\hat{\textbf{H}}_{s'})|}{\max(len(\hat{\textbf{H}}_{s}), len(\hat{\textbf{H}}_{s'}))}
    \label{eq:distance}
\end{equation}
where $Hamming(\hat{\textbf{H}}_{s'}, \hat{\textbf{H}}_{s})$ computes the normal Hamming distance between the common segments in $\hat{\textbf{H}}_{s}$ and $\hat{\textbf{H}}_{s'}$. 
To include the execution time of scenarios into Equation \ref{eq:distance},
we divide the maximum length of $\hat{\textbf{H}}_{s}$ and $\hat{\textbf{H}}_{s'}$ to obtain a normalized distance in $[0, 1]$. As shown in Figure \ref{fig:method_div}, the designed Hamming distance (e.g., 0.67 and 0.72) can represent the behavior differences in the three scenarios.

\subsubsection{Violation Feedback \label{method:sec_violation}}
Except for the behavior diversity, the primary target of ADS testing is to search for scenarios violating the given specifications.
Hence, in this paper, we also assess how far the scenario is from violating the specifications. 
As described in Section \ref{sec_ads_violation}, we mainly consider three types of specification violations: $collision$, $lines$, and  $destination$.
The violation degree of a seed $s$ is defined as follows:
\begin{equation}
    O = \sum_{v \in\{collision, lines, destination\}}f_{v}(s)
\end{equation}
where $f_{v}$ is the violation degree with respect to the corresponding specification.
The lower the value of $O$, the better the mutant. The violation degree for each specification is defined as follows:

\noindent \textit{Collision.} 
As given in Equation \ref{eqn:collision}, the criterion of collision is that the minimum distance between the ego vehicle and NPC vehicles in the scenario is equal to 0. 
Therefore, given the observation trace $\textbf{X}_s$ of a seed $s$, we define the collision degree as:
\begin{equation}
    f_{collision}(s) = D_b(p_0, s)
\end{equation}

\noindent \textit{Hitting Illegal Lines.} 
According to Equation \ref{eqn:rule}, the violation degree of not hitting illegal lines  $f_{lines}$ is calculated as:
\begin{equation}
    f_{lines}(s) = D_l(p_0, s).
\end{equation}

\noindent \textit{Reach Destination.} The violation degree of reaching the destination is defined by:
\begin{equation}
    f_{destination}(s) = \max(10-D_{p2p}(p_{0}^{t_{sim}}, p_{dest}), 0)
\end{equation}
where $D_{p2p}(p_{0}^{t_{sim}}, p_{dest})$ calculates the distance between the last position of the ego vehicle (i.e., $p_{0}^{t_{sim}}$) and the destination $p_{dest}$. 

\subsection{Energy Mechanism \label{sec_method:energy_mechanism}}
\tool considers both diversity and violation feedback, which aims to discover violations in diverse scenarios. 
However, the diversity feedback may compromise the detection of violations. 
In general, as violation scenarios are raw in the scenario space, the diversity feedback will add various scenarios to the input corpus, which are hard to trigger violations. Consequently, there is a high possibility of selecting seeds that cannot generate violations.
To mitigate this challenge, we propose an energy mechanism for seed selection and mutation, e.g., perform larger mutation for seeds that are difficult to trigger violations (with low energy).

\subsubsection{Energy Calculation}
\tool needs to calculate the energy in three cases: assign energy for the initial seeds (line 4 in Alogrithm~\ref{algo:BeDivFuzzer}), update the energy of the existing seed (line 19 in Algorithm~\ref{algo:BeDivFuzzer}) and assign energy for the newly generated seed (line 16 in Algorithm~\ref{algo:BeDivFuzzer}). For the first case, when the fuzzer starts, all initial seeds are assigned with the same energy (i.e., 1). Next, we will introduce the details about the other two cases, i.e., update the energy $E_s$ of $s$ and calculate the energy $E_{s'}$ for $s'$ during fuzzing (suppose $s'$ that is mutated from $s$ ).

We measure the violation potential of the seed $s$ from three factors: 1)  \textit{Failure Frequency}, i.e., how many violations have been discovered based on the seed $s$? 2) \textit{Violation Degree Change}, i.e., whether the new mutant $s'$ has a smaller violation degree than $s$, and 3) \textit{Selection Frequency}, i.e., the number of times the seed $s$ has been selected. Finally, the energy of $s$ is updated as:
\begin{eqnarray}\label{eqn:energy}
     E_{s} = E_{s} + w_1 \cdot \Delta{E_F} +  w_2 \cdot \Delta{E_V} + w_3 \cdot  \Delta{E_S}
\end{eqnarray}
where $ \Delta{E_F} $, $\Delta{E_V}$ and $\Delta{E_S}$ represent the failure frequency, the violation change and the selection frequency, respectively. $w_1$, $w_2$ and $w_3$ are parameters that can adjust the weights of the three factors. 
\begin{itemize}[leftmargin=*]
    \item \textit{Failure Frequency}. The basic idea is that we do not want to spend too much time on a seed from which it is hard to discover violations.  
    For the seed $s$, we record the number of failed/non-failed test cases (denoted as \#F and \#NF) that are mutated from $s$. Then we calculate the corresponding energy update as follows:
\begin{equation}
\label{eq:failfreq}
\Delta{E_{F}} = \left\{
\begin{aligned}
{\#F}/{(\#NF + \#F)} & , & {s'} \text{ is a failed test case}, \\
-{\#NF \cdot 0.1}/{(\#NF + \#F)} & , &  {s'} \text{ is a benign test case}.
\end{aligned}
\right.
\end{equation}
Specifically, if the current mutant $s'$ is a failed case, we add a positive energy for $s$. Otherwise, a small negative energy is added to $s$. Note that a smaller weight (0.1) used in the benign case is to avoid deducting too much energy.

\item\textit{Violation Degree Change}. As it may be hard to  generate failure cases directly, we consider the violation degree change between $s$ and $s'$:
\begin{equation}
\label{eq:viodegree}
\Delta{E_{V}} =  \frac{O_{{s}} - O_{{s'}}}{1-dis(\hat{\textbf{H}}_{s'},\hat{\textbf{H}}_{s}) + \gamma},
\end{equation}
where $\gamma$ is a small value (e.g., $10^{-5}$) to avoid zero denominator.
Intuitively, if $s'$ gets a lower violation degree than $s$ that meets our expectation, reward energy is added to $s$.  Otherwise, penalty energy is added to $s$. In addition, the diversity is considered (the denominator in Equation~\ref{eq:viodegree}), i.e., more energy is rewarded if $s'$ and $s$ have more different behavior.
\item \textit{Selection Frequency.} To avoid the fuzzer selecting a seed many times, which may lead to local convergence,  we reduce the energy of the seed when it is selected. In this paper, we select a fixed energy decay value, i.e., $ \Delta{E_{S}} = -0.05$.
\end{itemize}

In addition, we assign the initial energy for the newly generated mutant $s'$. Different from the initial seed with a fixed initial energy (i.e., 1), we calculate the initial energy of $s'$ as:
\begin{eqnarray}
     E_{s'} = 1 + w_2 * \Delta{E_V}
\end{eqnarray}
where $\Delta{E_V}$ is the change of the violation degree (see Equation~\ref{eq:viodegree}) .

\subsubsection{Energy-Based Seed Selection}
Based on the energy, we design a seed selection strategy. For each seed $s_i$, we calculate a selection probability as:
\begin{equation}
    p_{i} = \frac{\max (E_{s_i}, 0)}{\sum_{s\in \textbf{Q}}{E_{s}}}.
\end{equation}

We use the $\max$ function to make sure all probabilities are non-negative. The seed with higher energy will have a higher probability to be selected for mutation.

\begin{figure}[!t]
    \centering
    \includegraphics[width=\linewidth]{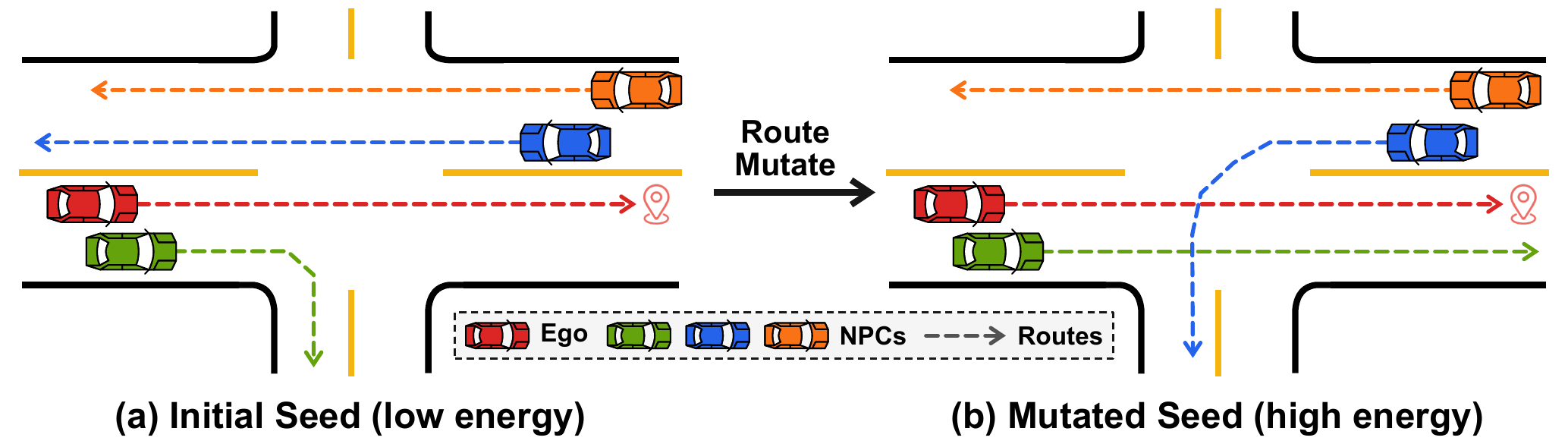}
    \caption{Illustration of Route Mutation.}
    \label{fig:method_mutation}
\end{figure}

\subsubsection{Energy-Based Adaptive Mutation}
Except for the seed selection, we also propose an adaptive mutation strategy that adopts different mutation methods for seeds with different levels of energy. Basically, the mutation is to modify the behaviors of NPC vehicles such that the violation happens to the ego vehicle. Specifically, for an NPC, we can change its route or only the waypoints. For the seed with the lower energy that is not easy to cause violations, we make major changes to the routes of some NPCs such that it has a significant difference from its parent. 
For example, as shown in Figure \ref{fig:method_mutation}(a), only mutating the waypoints of NPCs in the initial seed cannot generate critical scenarios, so it has lower energy based on our computation and requires changing the routes of the NPCs,  as shown in Figure \ref{fig:method_mutation}(b).
For the seed with higher energy, which means it is  more likely to cause violations, we make minor changes (i.e., the waypoints) without modifying the routes of NPC vehicles.

Algorithm \ref{algo:adaptive_mutation} shows the adaptive mutation. It performs the waypoint mutation if the seed has higher energy (line 4). Otherwise, the route mutation is performed (line 6). For the route mutation, it has a probability (line 9) to select and mutate an NPC $v$ by changing its route. The function RouteSample randomly samples a new route $v_r$ for the selected NPC (line 10). The function UniformSample uniformly samples waypoints $v_w$ for the newly generated route $v_r$ (line 11). Finally, a mutant $s'$ is generated (line 12). For the waypoint mutation that does not change the route of an NPC, we only change the waypoints (e.g., the velocity and the position) (line 16).

\begin{algorithm}[!t]
\small
    \SetKwInOut{Input}{Input}
    \SetKwInOut{Output}{Output}
    \SetKwInOut{Para}{Parameter}
    \SetKwComment{Comment}{/*}{*/}
    \SetKwFunction{MutationRoute}{{\textbf{MutationRoute}}}
    \SetKwFunction{MutationWaypoint}{{\textbf{MutationWaypoint}}}
    \SetKwFunction{AdaptiveMutation}{{\textbf{AdaptiveMutation}}}

\SetKwFunction{GetNPC}{GetAllNPC}
    \SetKwFunction{UniformRandom}{UniformRandom}
    \SetKwFunction{UpdateScenario}{UpdateScenario}
    \SetKwFunction{RandomSample}{RouteSample}
    \SetKwFunction{UniformSample}{UniformSample}
    \SetKwFunction{GaussMutation}{GaussMutation}
    \SetKwProg{Fn}{Function}{:}{}
    
    \Input{Selected scenario seed $s$ with energy $E_{s}$\\
    Energy threshold $\theta{e}$
    }
    \Output{Mutated scenario seed $s'$}
     \Para{Random mutation threshold $\epsilon$ 
           }
 \Fn{\AdaptiveMutation{$s$, $E_{S}$}}{
        $V \gets \GetNPC (s)$ \\
        \eIf{$E_s > \theta_{e}$}{
            \Comment*[l]{Mutate waypoints for high-energy seeds} 
            \Return $\MutationWaypoint(s, V)$
        }{
            \Comment*[l]{Mutate routes for low-energy seeds} 
            \Return $\MutationRoute(s, V)$
        }
    }
    \Fn{\MutationRoute{$s, V$}}{
        \For{$v \in V$} 
        {
            \If{$\UniformRandom()>\epsilon$}{
                \Comment*[l]{Sample a new route randomly and its waypoints uniformly for $v$} 
                $v_r \gets \RandomSample()$ \\
                $v_w \gets \UniformSample(v_r)$ 
            }
        }
        \Return $s' \gets \UpdateScenario(s)$
    }
    \Fn{\MutationWaypoint{$s, V$}}{
        \For{$v \in V$} 
        {
            \If{$\UniformRandom()>\epsilon$}{ 
                $v_w \gets \GaussMutation(v_w)$ \Comment*[l]{Mutate waypoints}
            }
        }
        \Return $s' \gets \UpdateScenario(s)$
    }
\caption{Energy-Based Adaptive Mutation}
\label{algo:adaptive_mutation}
\end{algorithm}

\section{Empirical Evaluation}
In this section, we evaluate the effectiveness of \tool in generating diverse and critical scenarios. In particular, we will answer the following research questions:

\noindent \textbf{RQ1}: Can \tool effectively find diverse violations for ADSs compared to the selected baselines?

\noindent \textbf{RQ2}: How useful are \miner and energy mechanism in improving \tool in terms of generating diverse and critical scenarios?

\noindent \textbf{RQ3}: How does the temporal feature affect the testing results? 

To answer the three research questions, we implemented \tool in the simulation environment built with 
 Baidu Apollo 6.0~\cite{apollo} (the ADS under test) and LGSVL 2021.3~\cite{rong2020lgsvl}.
We run \tool on the following four functional scenarios related to intersections and roads in the real world as they are very representative driving tasks in the real world and have been widely tested in the literature~\cite{thorn2018framework,av_fuzzer,althoff2018automatic,gambi2019generating}.

\noindent \textbf{S1}: Ego \textbf{goes straight} through a non-signalized intersection.

\noindent \textbf{S2}: Ego \textbf{turns left} at a non-signalized intersection.

\noindent \textbf{S3}: Ego \textbf{follows a lane} at a straight road with four lanes.

\noindent \textbf{S4}: Ego \textbf{changes lanes} at a straight road with four lanes.

We compare \tool with three representative techniques: random testing, AVFuzzer \cite{av_fuzzer}, and SAMOTA \cite{icse_samota}.
Random testing does not have any feedback. It randomly selects a test case from the seed corpus and mutates it. 
AVFuzzer is a two-phase method where the genetic algorithm is applied to search for scenarios with a high risk of violating safety requirements, and a local fuzzer is applied to search for safety violations based on the generated high-risk scenarios.  
SAMOTA is the state-of-the-art that extends the existing many-objective search algorithms with surrogate models for efficiently finding safety-related violations. 
Note that SAMOTA is originally evaluated on the Advanced Driver Assistance Systems (ADAS) Pylot and the simulator CARLA. 
For comparison, we customize it to our simulation environment, i.e., Apollo+LGSVL. 

\noindent \textbf{Configuration.} The cluster number of \miner is 10, which has a better computing efficiency trade-off. 
In our experiments, we set the time interval as 1s for mining ADS behavior states. 
Specifically, the aligned sampling frequency is 0.1 s, and the sliding window size is set to 10. 
In this way, we can extract the temporal features of the ego vehicle within each second. 
Similar to \cite{av_fuzzer}, the thresholds $\theta_{c}$ and $\theta_{e}$ in Algorithm \ref{algo:BeDivFuzzer}, and $\epsilon$ in Algorithm \ref{algo:adaptive_mutation} are empirically set to 0.4, 0.5 and 0.5, respectively. Specifically, the behavior guidance threshold $\theta_{c}$ is selected based on some sampled scenarios with different ego behaviors, e.g., the three scenarios in Figure \ref{fig:method_div}.
The weights $w_{1}$, $w_{2}$ and $w_{3}$ in Equation \ref{eqn:energy} are empirically set to 0.5, 0.5, and 1, respectively. For comparison, we run \tool and alternatives on each scenario.
In case of randomness, each method is repeated 5 times for each scenario, resulting in $5 \times 4 \times 4 = 80$ executions.
Each execution lasts for 12 hours.
All results and implementation details can be found on our website~\cite{ourweb}.

\noindent \textbf{Metrics.} In the experiments, we mainly select two metrics to measure the effectiveness: the number of violations detected and the number of unique violations. Specifically, for each method, we first collect the corresponding violated scenarios. Then, we manually analyze the violated scenarios and classify them into three groups based on the specifications they violate. To evaluate the diversity of generated violations, we manually inspect the generated violated scenarios of all methods (baselines and \tool) and get abstract violations in terms of ego behavior and the reasons for violations. Finally, we summarize the unique violation patterns that are used to measure the number of unique violations.

\subsection{RQ1: Results of Discovering Violations \label{sec:rq1}}
\subsubsection{Total Violations Results}
Table \ref{tab:RQ1} shows the results of total violations and unique violations found by \tool and other baselines.
\textbf{S1}, \textbf{S2}, \textbf{S3} and \textbf{S4} represent the four functional scenarios. \textbf{R1}, \textbf{R2} and \textbf{R3} show the violations with different specifications, i.e., collision, hitting lines, and reaching the destination, respectively. 
From the results, we can find that on average, \tool can discover more violated scenarios than others on each specification in 12 hours. For intersection scenarios (i.e., \textbf{S1} and \textbf{S2}), \tool performs better than the best of other methods (AVFuzzer) on all specifications: \textbf{R1} (14.6 vs. 13.6), \textbf{R2} (12.4 vs.10.0) and \textbf{R3} (36.6 vs. 18.0). In the straight-road scenarios, i.e., \textbf{S3} and \textbf{S4}, \tool also achieve satisfying performance on finding violations. 
Compared with the best of the existing methods (AVFuzzer), \tool improves the number of violated scenarios significantly on \textbf{R3} (improve from 21.2 to 38.2 in \textbf{S3}) and achieves competitive results on both \textbf{R1} and \textbf{R2}. In summary, the quantity results in Table \ref{tab:RQ1} show that \tool can effectively find more violations in a given time. 
The main reason is that \tool uses the scenario diversity and violation degree to guide the generation of violated scenarios. 
However, in AVFuzzer, the fitness function is only used to search for scenarios that have high violation potential, and a local fuzzer is used to search for violated scenarios, which is not efficient enough.  With behavior diversity, \tool could discover more diverse scenarios that can lead to violated scenarios, which can mitigate the local optimal of guided searching methods.

\begin{table}[t!]
    \centering
    \caption{Comparison results with baselines. \textit{Total/Unique Violation} is the average of the total/unique number of violations discovered in 5 executions. Results in \textbf{bold} denote the best.}
    \resizebox{\linewidth}{!}{
        \begin{tabular}{clcccccccc}
            \toprule[1pt]
            \multirow{2} * {\textbf{Scenario}} & \multirow{2} * {\textbf{Method}} & \multicolumn{4}{c}{\textbf{Total Violation}} & \multicolumn{4}{c}{\textbf{Unique Violation}} \\
            \cmidrule(lr){3-6}\cmidrule(lr){7-10} 
             & & \textbf{R1} & \textbf{R2} & \textbf{R3} & \cellcolor{lightgray!30}\textbf{Sum} & \textbf{R1} & \textbf{R2} & \textbf{R3} & \cellcolor{lightgray!30}\textbf{Sum} \\
            \midrule
            \multirow{4} * {\textbf{S1}} & \textbf{Random} & 11.2 & 10.0 & 16.6 & \cellcolor{lightgray!30}37.8 & 3.0 & 0.8 & 2.4 & \cellcolor{lightgray!30}6.2 \\
             & \textbf{AVFuzzer} & 13.6 & 2.4 & 17.8 & \cellcolor{lightgray!30}33.8 & 3.6 & 1.4 & 1.8 & \cellcolor{lightgray!30}6.8 \\
             & \textbf{SAMOTA} & 9.2 & 0.0 & 18.0 & \cellcolor{lightgray!30}27.2 & 2.6 & 0.0 & 1.6 & \cellcolor{lightgray!30}4.2 \\
             & \textbf{BehAVExplor} & \textbf{14.6} & \textbf{12.4} & \textbf{36.6} & \cellcolor{lightgray!30}\textbf{63.6} & \textbf{4.4} & \textbf{2.2} & \textbf{2.8} & \cellcolor{lightgray!30}\textbf{9.4} \\
            \midrule
            \multirow{4} * {\textbf{S2}} & \textbf{Random} & 3.0 & 7.6 & 23.8 & \cellcolor{lightgray!30}34.4 & 1.8 & 1.6 & 1.4 & \cellcolor{lightgray!30}4.8 \\
             & \textbf{AVFuzzer} & 4.2 & 18.4 & 25.8 & \cellcolor{lightgray!30}48.4 & 2.6 & 1.6 & 1.6 & \cellcolor{lightgray!30}5.8 \\
             & \textbf{SAMOTA} & 3.0 & 14.2 & 21.6 & \cellcolor{lightgray!30}38.8 & 1.4 & 1.6 & 1.4 & \cellcolor{lightgray!30}4.4\\
             & \textbf{BehAVExplor} & \textbf{9.6} & \textbf{27.0} & \textbf{41.4} & \cellcolor{lightgray!30}\textbf{78.0} & \textbf{3.8} & \textbf{1.8} & \textbf{2.4} & \cellcolor{lightgray!30}\textbf{8.0} \\
            \midrule
            \multirow{4} * {\textbf{S3}} & \textbf{Random} & 2.2 & 3.0 & 10.4 & \cellcolor{lightgray!30}15.6 & 0.8 & 1.4 & 2.2 & \cellcolor{lightgray!30}4.4  \\
             & \textbf{AVFuzzer} & 2.6 & 6.0 & 21.2 & \cellcolor{lightgray!30}29.8 & 1.0 & 0.8 & 1.8 & \cellcolor{lightgray!30}3.6 \\
             & \textbf{SAMOTA} & 0.0 & 0.4 & 13.2 & \cellcolor{lightgray!30}13.6 & 0.0 & 0.2 & 1.6 & \cellcolor{lightgray!30}1.8 \\
             & \textbf{BehAVExplor} & \textbf{4.2} & \textbf{7.8} & \textbf{38.2} & \cellcolor{lightgray!30}\textbf{50.2} & \textbf{1.8} & \textbf{1.6} & \textbf{2.2} & \cellcolor{lightgray!30}\textbf{5.6} \\
            \midrule
            \multirow{4} * {\textbf{S4}} & \textbf{Random} & 5.8 & \textbf{9.0} & 23.8 & \cellcolor{lightgray!30}38.6 & 1.4 & 1.2 & 2.2 & \cellcolor{lightgray!30}4.8 \\
             & \textbf{AVFuzzer} & 6.2 & 6.2 & 32.2 & \cellcolor{lightgray!30}44.6 & 1.6 & \textbf{1.4} & 2.2 & \cellcolor{lightgray!30}5.2 \\
             & \textbf{SAMOTA} & 0.6 & 2.0 & 11.0 & \cellcolor{lightgray!30}13.6 & 0.6 & 1.0 & 1.8 & \cellcolor{lightgray!30}3.4 \\
             & \textbf{BehAVExplor} & \textbf{6.2} & 6.6 & \textbf{38.8} & \cellcolor{lightgray!30}\textbf{61.6} & \textbf{1.8} & 1.0 & \textbf{2.8} & \cellcolor{lightgray!30}\textbf{5.6}\\
            \bottomrule[1pt]
        \end{tabular}
    }
    \label{tab:RQ1}
\end{table}

\begin{figure*}[t]
    \centering
    \includegraphics[width=\linewidth]{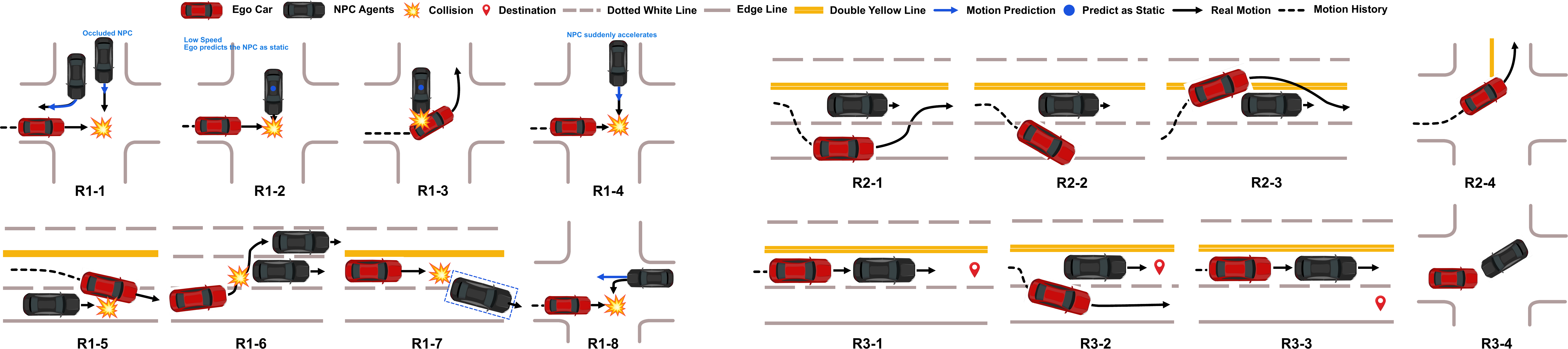}
    \caption{Unique violations (detailed at Section \ref{sec:rq1}). The red vehicle is the ego, and the black vehicles are NPC.} 
    \label{fig:RQ1-ex}
\end{figure*}

\subsubsection{Unique Violation Classification}
To verify the diversity of generated violations, we manually classify the violations discovered by \tool and the three baselines. The classification results are shown in Figure \ref{fig:RQ1-ex}, which contains all unique violations discovered by \tool.
Note that some issues may not be covered by baselines. 

 \textbf{Unique Violation of Collision (\textbf{R1}).} The left part of Figure \ref{fig:RQ1-ex} shows the violations against collision, which contains the following 8 unique violations:

\noindent \textbf{R1-1:} No prediction of partially occluded vehicles (Figure \ref{fig:RQ1-ex} (R1-1)). When an NPC vehicle partially occludes another one, Apollo can not predict the motion of the occluded vehicle exactly, resulting in a collision between the ego and the occluded vehicles. 

\noindent \textbf{R1-2:} No motion prediction of low-speed vehicles (Figure \ref{fig:RQ1-ex} (R1-2)). Apollo regards a low-speed NPC vehicle as a stationary one and has no motion prediction. Hence, Apollo makes a wrong decision and causes a collision with the low-speed vehicle.

\noindent \textbf{R1-3:} Collision with a temporary stopping vehicle (Figure \ref{fig:RQ1-ex} (R1-3)). When the ego vehicle is turning left, and an NPC vehicle is approaching, Apollo will make a stop decision temporarily. However, the ego vehicle restarts to move quickly as Apollo makes another overtaking decision.
Hence, a collision occurs.

\noindent \textbf{R1-4:} Collision with a suddenly changed vehicle (Figure \ref{fig:RQ1-ex} (R1-4)). Apollo does not perform full braking and causes a collision when an NPC vehicle changes its behavior suddenly (e.g., rapid acceleration). %or hard braking

\noindent \textbf{R1-5:} Collision in an aggressive lane-changing behavior (Figure \ref{fig:RQ1-ex} (R1-5)). Apollo sometimes does not decelerate to wait for other close vehicles moving on the target lane during lane changing or cut-in, which finally results in a collision.

\noindent \textbf{R1-6:} Rear-end collision during lane changing (Figure \ref{fig:RQ1-ex} (R1-6)). Apollo does not stop lane changing in time when an NPC is detected on the target lane nearby.
Therefore, a collision occurs. 

\noindent \textbf{R1-7:} Rear-end collision with large vehicles, e.g., school bus (Figure \ref{fig:RQ1-ex} (R1-7)). When a large NPC vehicle makes a turn, it may occupy the adjacent lane. 
However, Apollo cannot detect such an occupation and causes a collision with the large vehicle.

\noindent \textbf{R1-8:} Wrong motion prediction of NPC vehicles (Figure \ref{fig:RQ1-ex} (R1-8)).
In the case of multiple routes for an NPC vehicle, Apollo sometimes makes a wrong prediction, e.g., predicting a left turn as a straight motion.
Thus, Apollo does not slow down and causes a collision. 

\textbf{Unique Violation of Hitting Illegal Lines (\textbf{R2}).} 
\tool discovers four unique violations against \textbf{R2}.
The diagrams of the four violations are given at the right top of Figure \ref{fig:RQ1-ex}.

\noindent \textbf{R2-1:} Overtaking with slow heading recovery (Figure \ref{fig:RQ1-ex} (R2-1)). 
Apollo sends a large steer command to the ego vehicle for overtaking but computes a relatively small steer command to recover the ego vehicle's heading. Hence, the ego vehicle deviates from the center of the target land and moves along the edge line of the lane.

\noindent \textbf{R2-2:} Aggressive overtaking (Figure \ref{fig:RQ1-ex} (R2-2)). When the ego vehicle performs a sharp turn with rapid acceleration during lane changing or overtaking, it will hit the edge line and may be stuck forever between two adjacent lanes.

\noindent \textbf{R2-3:} Wrong overtaking direction (Figure \ref{fig:RQ1-ex} (R2-3)). In this issue, the ego vehicle executes overtaking by moving along the lane in an opposite direction, while the adjacent lane in the same direction is empty. Such behavior will hit illegal lines.

\noindent \textbf{R2-4:} Sharp turning. In a left-turn scenario (Figure \ref{fig:RQ1-ex} (R2-4)), if the speed of the ego vehicle is slow and the steering angle is relatively large, the ego vehicle will hit the yellow line and continue to press the yellow line after left turn.

\textbf{Unique Violation of Reaching Destination (\textbf{R3}).} The scenarios violated \textbf{R3} are classified into four different unique violations (the right bottom in Figure \ref{fig:RQ1-ex}). 

\noindent \textbf{R3-1:} No overtaking action (Figure \ref{fig:RQ1-ex} (R3-1)). In this situation, the ego vehicle follows a slow-speed vehicle ahead, rather than overtaking, resulting in the ego vehicle cannot arrive at the destination within the given time duration.

\noindent \textbf{R3-2:} Wrong overtaking decision (Figure \ref{fig:RQ1-ex} (R3-2)).
The ego vehicle performs overtaking when an NPC vehicle ahead, which is ahead of the ego vehicle's destination in the same lane.
Thus, the ego vehicle cannot reach its destination.

\noindent \textbf{R3-3:} Late lane changing (Figure \ref{fig:RQ1-ex} (R3-3)). In the lane-changing scenario, Apollo chooses to follow a low-speed NPC vehicle and does not take lane-changing action early, regardless of whether there are NPC vehicles in the target lane. This causes the ego not to reach the destination on time but to plan a longer route.

\noindent \textbf{R3-4:} Deadlock with NPC vehicles (Figure \ref{fig:RQ1-ex} (R3-4)).
When the ego and NPC vehicles move to a conflicting area simultaneously, there may be deadlocks between the ego vehicle and NPC vehicles as 
Apollo cannot detect such situations in advance.

\subsubsection{Diversity of Violations}
We summarize the number of unique violations discovered by the four methods in their executions. 
From the results shown in Table \ref{tab:RQ1}, we can find on average, \tool can discover 9.4, 8.0, 5.6, and 5.6 unique violations against the three specifications in \textbf{S1}, \textbf{S2}, \textbf{S3}, and \textbf{S4}, respectively. 
While the best average results of other methods are 6.8 (AVFuzzer), 5.8 (AVFuzzer), 4.4 (Random), and 5.2 (AVFuzzer) unique violations in \textbf{S1}, \textbf{S2}, \textbf{S3}, and \textbf{S4}, respectively. 
Therefore, \tool improves the number of discovered unique violations by 38.2\%, 37.9\%, 27.3\%, and 7.7\% in \textbf{S1}, \textbf{S2}, \textbf{S3}, and \textbf{S4}, respectively. We note that \tool performs worse than others on finding \textbf{R2} on scenario \textbf{S4}, which indicates limited improvement on simple scenarios. But to sum up, our \tool is able to find more unique violations as well as total violations.
Table~\ref{tab:RQ1-div} lists the total number of unique violations discovered in all 5 repetitions.
The results show that \tool can discover the most number of unique violations and cover all unique violations discovered by baselines.
The violation examples can be found on our website in~\cite{ourweb}.

\begin{table}[!t]
    \centering
    \caption{Results of the total number of unique violations.}
    \small
    \resizebox{\linewidth}{!}{
        \begin{tabular}{lcccc|c}
        \toprule[1pt]
        \textbf{Method} & \textbf{R1} & \textbf{R2} & \textbf{R3} & \textbf{Sum} & \textbf{Details} \\
        \midrule
        \textbf{Random} & 5 & 3 & 2 & {10} & R1-2, R1-4, R1-6 to R1-8, R2-2 to R2-4, R3-1, R3-3 \\
        \textbf{AVFuzzer} & 6 & 3 & 3 & {12} & R1-1, R1-2, R1-4, R1-6 to R1-8, R2-2 to R2-4, R3-1, R3-3, R3-4 \\
        \textbf{SAMOTA} & 4 & 2 & 2 & {8} & R1-2, R1-4, R1-7, R1-8, R2-2, R2-4, R3-1, R3-3\\
        \textbf{\tool} & \textbf{8} & \textbf{4} & \textbf{4} & \textbf{16} & R1-1 to R1-8, R2-1 to R2-4, R3-1 to R3-4\\
        \bottomrule[1pt]
        \end{tabular}
    }
    % \caption{Results of the total number of unique violations.}
    \label{tab:RQ1-div}
\end{table}
\begin{table}[!t]
    \centering
    \caption{Comparison of performance characteristics.}
    \small
    \resizebox{0.9\linewidth}{!}{
        \begin{tabular}{ccccc}
        \toprule[1pt]
        \textbf{\makecell[c]{Time (hour)}} & \textbf{Random} & \textbf{AVFuzzer} & \textbf{SAMOTA} & \textbf{BehAVExplor} \\
        \midrule
        \textbf{Execution} & 11.30 & 10.65 & 6.26 & 10.82 \\
        \textbf{Analysis} & 0.70 & 1.35 & 5.74 & 1.18 \\
        \bottomrule[1pt]
        \end{tabular}
    }
    % \caption{Comparison of performance characteristics.}
    \label{tab:RQ4-time}
\end{table}

In addition, we studied the efficiency of each tool. The testing time on a scenario (12 hours) can be split into two parts: execution time and analysis time. Execution time is the time taken to run the generated scenarios, while analysis time includes other processes such as seed selection, mutation, feedback collection, and other logging processes. We calculate the average execution and analysis time for all scenarios. Table~\ref{tab:RQ4-time} displays the results in detail. Notably, Random has the shortest analysis time, as it does not have any guidance. Comparing the results of AVFuzzer and SAMOTA with \tool, we can observe that \tool has less analysis time and more execution time, demonstrating the high efficiency of our guidance (e.g., behavior diversity analysis). The analysis time of AVFuzzer is prolonged due to the time-consuming seed generation process during restarts, while SAMOTA's analysis phase is hindered by the extensive training and inference of surrogate models.

\begin{center}
\fcolorbox{black}{gray!10}{\parbox{0.96\linewidth}{\textbf{Answer to RQ1: Compared with the existing methods, \tool not only can discover more violated scenarios but also can generate more diverse violations.}}}
\end{center}

\subsection{RQ2: Ablation Study}
\noindent \textbf{Evaluation Setup.} 
We design an ablation study to investigate the usefulness of the behavior guidance \miner and the energy mechanism. 
Specifically, we choose Random as our baseline, where  the behavior guidance and the energy mechanism in BehAVExplor are replaced by random selections. 
To evaluate the usefulness of \miner and the energy mechanism, we create two variants by removing \miner (denoted as \textbf{w/o BG}) and the energy mechanism  (denoted as \textbf{w/o Energy}) from \tool, respectively. 
We evaluate both variants using the same set of scenarios (\textbf{S1} to \textbf{S4}) and settings as described in Section \ref{sec:rq1}, where each scenario is executed for 12 hours and repeated 5 times. We compare their results  with Random and \tool in terms of the number of total violations and the number of unique violations.

\textbf{Usefulness of Behavior Guidance (BG).}
The design of behavior guidance aims to generate scenarios with more diverse behaviors so as to discover more unique violations. 
From the last four columns of Table \ref{tab:RQ2_RQ3}, we can find that \textbf{\tool} can discover more unique violations than \textbf{w/o BG} in each scenario in terms of the total number of unique violations. Thus, we can conclude that the behavior guidance (BG) is useful in finding diverse violations.

\textbf{Potential Conflicts between Diversity and Violations.}
We can also find that \textbf{w/o BG} detects more violations than \tool (see Total Violation), which indicates the  potential conflict between diversity and violation detection. Specifically, when diversity is increased, the possibility of detecting violations may be reduced. We further analyzed the correlation between diversity and violation detection by calculating Pearson Correlation Coefficient $r_{BF}$ \cite{official_pearson, web_pearson} between the violation fitness and the behavior distance on the results of \tool. The results are shown in Table \ref{tab:r_BF}. We can see that the correlation value $r_{BF}$ is located in the interval $[-0.3, 0]$ with $p$-$value < 0.001$ for all scenarios, which shows a negative correlation between the violation detection and diversity.

\begin{table}[!t]
    \centering
    \caption{
    Results of ablation study. Results in \textbf{bold} denote the best, while the underlined ones indicate the second best.
    }
    %\vspace{-0.2cm}
    \resizebox{\linewidth}{!}{
        \begin{tabular}{clcccccccc}
        \toprule[1pt]
        \multirow{2} * {\textbf{Scenario}} & \multirow{2} * {\textbf{Method}} & \multicolumn{4}{c}{\textbf{Total Violation}} & \multicolumn{4}{c}{\textbf{Unique Violation}} \\
        \cmidrule(lr){3-6}\cmidrule(lr){7-10} 
         &  & \textbf{R1} & \textbf{R2}  & \textbf{R3} & \cellcolor{lightgray!30}\textbf{Sum} & \textbf{R1} & \textbf{R2} & \textbf{R3} & \cellcolor{lightgray!30}{\textbf{Sum}}\\
        \midrule
        \multirow{4} * {\textbf{S1}} & \textbf{Random} & 11.2 & 10.0 & 16.6 & \cellcolor{lightgray!30}37.8 & 3.0 & 0.8 & 2.4 & \cellcolor{lightgray!30}6.2 \\
        & \textbf{w/o BG} & \textbf{15.2} & \textbf{14.0} & \textbf{39.6} & \cellcolor{lightgray!30}\textbf{68.8} & 2.8 & 2.2 & 1.0 & \cellcolor{lightgray!30}6.0 \\
        & \textbf{w/o Energy} & 12.6 & 11.2 & 31.0 & \cellcolor{lightgray!30}45.4 & \underline{4.2} & \underline{2.0} & \underline{2.8} & \cellcolor{lightgray!30}\underline{8.8} \\
        & \textbf{BehAVExplor} & \underline{14.6} & \underline{12.4} & \underline{36.6} & \cellcolor{lightgray!30}\underline{63.6} & \textbf{4.4} & \textbf{2.2} & \textbf{2.8} & \cellcolor{lightgray!30}\textbf{9.4} \\
        \midrule
        \multirow{4} * {\textbf{S2}} & \textbf{Random} & 3.0 & 7.6 & 23.8 & \cellcolor{lightgray!30}34.4 & 1.8 & 1.6 & 1.4 & \cellcolor{lightgray!30}4.8 \\
        & \textbf{w/o BG} & 6.0 & \textbf{33.0} & \textbf{48.8} & \cellcolor{lightgray!30}\textbf{87.8} & 1.4 & 1.2 & 2.0 & \cellcolor{lightgray!30}4.6 \\
        & \textbf{w/o Energy} & \underline{8.0} & 21.0 & \underline{46.0} & \cellcolor{lightgray!30}75.0 & \underline{3.0} & \underline{1.6} & \underline{2.6} & \cellcolor{lightgray!30}\underline{7.2} \\
        & \textbf{BehAVExplor} & \textbf{9.6} & \underline{27.0} & 41.4 & \cellcolor{lightgray!30}\underline{78.0 } & \textbf{3.8} & \textbf{1.8} & \textbf{2.4} & \cellcolor{lightgray!30}\textbf{8.0} \\
        \midrule
        \multirow{4} * {\textbf{S3}} & \textbf{Random} & 2.2 & 3.0 & 10.4 & \cellcolor{lightgray!30}15.6 & 0.8 & 1.4 & 2.2 & \cellcolor{lightgray!30}4.4 \\
        & \textbf{w/o BG} & \textbf{4.8} & \textbf{8.4} & \textbf{45.6} & \cellcolor{lightgray!30}\textbf{58.8} & 1.0 & 1.0 & 2.0 & \cellcolor{lightgray!30}4.0 \\
        & \textbf{w/o Energy} & 3.4 & 3.8 & 32.2 & \cellcolor{lightgray!30}39.4 & \underline{1.6} & \underline{1.6} & \underline{2.2} & \cellcolor{lightgray!30}\underline{5.4}  \\
        & \textbf{BehAVExplor} & \underline{4.2} & \underline{7.8} & \underline{38.2} & \cellcolor{lightgray!30}\underline{50.2} & \textbf{1.8} & \textbf{1.6} & \textbf{2.2} & \cellcolor{lightgray!30}\textbf{5.6} \\
        \midrule
        \multirow{4} * {\textbf{S4}} & \textbf{Random} & 5.8 & 9.0 & 23.8 & \cellcolor{lightgray!30}38.6 & 1.4 & \textbf{1.2} & 2.2 & \cellcolor{lightgray!30}4.8 \\
        & \textbf{w/o BG} & \textbf{8.8} & \textbf{17.6} & \textbf{39.6} & \cellcolor{lightgray!30}\textbf{66.0} & 1.0 & 0.6 & 1.8 & \cellcolor{lightgray!30}3.4\\
        & \textbf{w/o Energy} & 3.0 & \underline{10.0} & 25.8 & \cellcolor{lightgray!30}38.8 & \textbf{2.0} & 1.0 & \underline{2.6} & \cellcolor{lightgray!30}\underline{5.6} \\
        & \textbf{BehAVExplor} & \underline{6.2} & 6.6 & \underline{38.8} & \cellcolor{lightgray!30}\underline{61.6} & \underline{1.8} & \underline{1.0} & \textbf{2.8} & \cellcolor{lightgray!30}\textbf{5.6} \\
        \bottomrule[1pt]
        \end{tabular}
    }
    % \caption{
    % Results of ablation study. Results in \textbf{bold} denote the best, while the underlined ones indicate the second best.
    % }
    \label{tab:RQ2_RQ3}
\end{table}

\begin{table}[!t]
    \centering
    \caption{Correlation between behavior distance and fitness.}
    \small
    \resizebox{0.9\linewidth}{!}{
        \begin{tabular}{ccccc}
        \toprule[1pt]
        \textbf{Metric} & \textbf{S1} & \textbf{S2} & \textbf{S3} & \textbf{S4} \\
        \midrule
        \textit{$r_{BF}$} & -0.22 & -0.23 & -0.23 & -0.28 \\
        \textit{p-value} & $9.87 \times 10^{-10}$ & $1.08 \times 10^{-9}$ & $1.21 \times 10^{-7}$ & $5.92 \times 10^{-11}$ \\
        \bottomrule[1pt]
        \end{tabular}
    }
    \label{tab:r_BF}
\end{table}

\textbf{Usefulness of Energy Mechanism (Energy).} 
To mitigate the potential conflicts between increasing diversity and detecting violations, we introduced the Energy Mechanism. Consider Column [Total Violation, Sum] in Table \ref{tab:RQ2_RQ3}, we can observe that the performance of the \textbf{w/o Energy} variant drops on each scenario compared to \textbf{\tool}, which confirms the usefulness of Energy in mitigating the conflicts, i.e., improving the total number of violations in \tool with behavior guidance integrated. 
However, while this approach is useful in improving violation detection to mitigate such conflicts, it is not always effective in every scenario due to its heuristic-based nature. There are some factors that can affect its effectiveness, such as randomness, scenario complexity, and the additional performance cost introduced by the Energy Mechanism.
For example, there are two unexpected results (S2-R3, S4-R2) in Total Violation where \textbf{w/o Energy} detects more violations than \tool (46.0-41.1, 10.0-6.6).
In the case of S2-R3, we conjecture that the search might have converged to the local optima as S2 is a relatively simple scenario. Diversity guidance has the potential to mitigate the local optima by introducing diverse behaviors. In the case of S4-R2, we found that \textbf{w/o Energy} performed better than \tool, likely due to the additional computational cost of the Energy Mechanism, which resulted in \textbf{w/o Energy} running more scenarios (30+) than \tool in the given time.
We also observed that the Energy Mechanism not only improves total violation detection but could also enhance the detection of unique violations, although this effect is not always guaranteed to happen. For example, we found that in most cases, the performance of the \textbf{w/o Energy} variant under the Unique Violation column drops compared to \textbf{\tool}. 

\begin{center}
%\vspace{-0.1cm}
\fcolorbox{black}{gray!10}{\parbox{0.96\linewidth}{\textbf{Answer to RQ2: Both \miner and the energy mechanism are useful.  \miner can improve the diversity of the identified violations, while the energy mechanism can improve the number of violations after the integration of behavior guidance.
}}}
\end{center}

\subsection{RQ3: Effectiveness of Temporal Feature}
\begin{figure}[!t]
    \centering
    \includegraphics[width=0.8\linewidth]{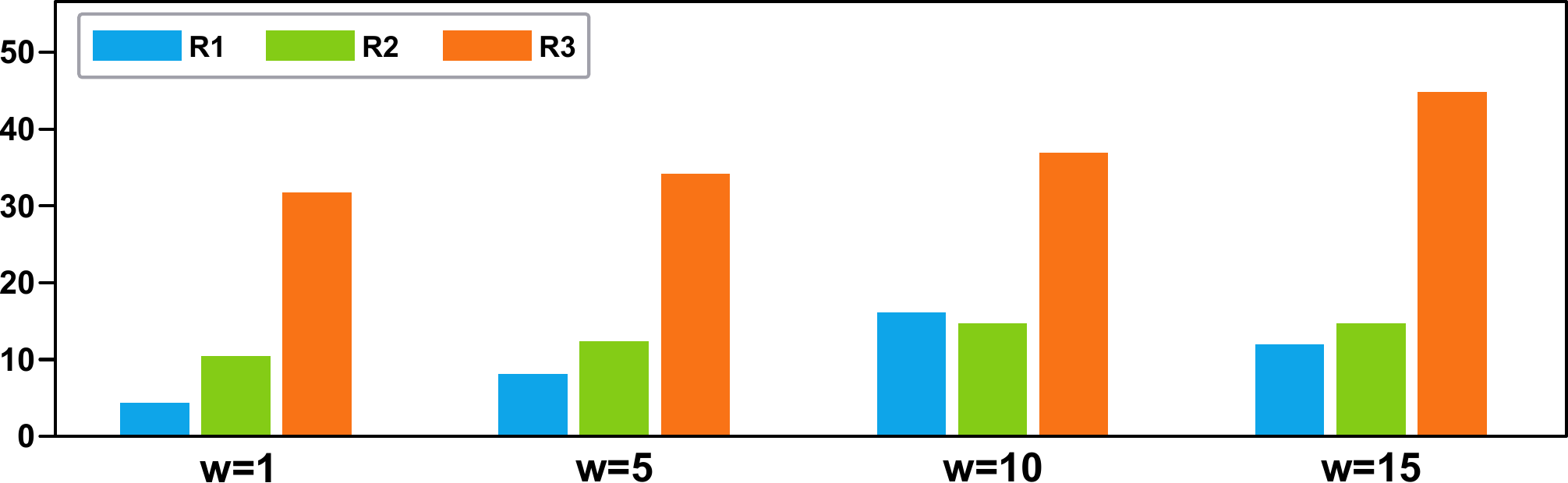}
    \caption{Comparison of different window settings (on S1).}
    \label{fig:rq2}
    % %\vspace{-0.6cm}
\end{figure}

As described in Section \ref{sec_method:behavior_miner}, as an important step to define behavior diversity, we adopt State Interpolation and State Merging in \miner to extract temporal features, which are used for behavior clustering. 
In this part, we evaluate the effectiveness of the temporal features, i.e., the effect of the sliding window. Specially, we set the window size $w$ as 1, 5, 10 and 15 to evaluate the different results. Note that $w=1$ means \tool does not consider the temporal features.  
Figure \ref{fig:rq2} shows the results with different window sizes. From the results, we can find that \tool with $w=1$ discovers the least number of violations.  With the increase of $w$, the total number of violations can generally get better. We also analyzed the diversity of the violations, the results have a similar trend, i.e., the larger the $w$, the diversity better. For example, the total numbers of unique violations are 5, 7, 10, and 8 when we set $w$ as 1, 5, 10, and 15, respectively. 

\begin{center}
%\vspace{-0.1cm}
\fcolorbox{black}{gray!10}{\parbox{0.96\linewidth}{\textbf{Answer to RQ3: Temporal feature extraction via sliding window can improve the effectiveness of violation discovering in \tool.}}}
\end{center}

\subsection{Threats to Validity}
\tool suffers from some threats due to the application of simulation-based testing. First, LGSVL suffers from the latency of LiDAR point cloud rendering, which results in the delay of the sensing data sent to the ADS's perception module and leads the ADS to make wrong decisions. 
We mitigate this threat by bypassing the perception module and sending the ground truth of the perception to the ADS, i.e., we mainly test the other modules except for perception.
Second, some modules in Apollo may suffer from high delays due to performance degradation after long-time continuous execution, which may result in failed scenarios and generate false-positive results by \tool. We implemented some solutions to mitigate this threat: (1) We used high-performance computers to conduct experiments; (2) During our experiments, we restarted the modules periodically to reduce module delays; (3) We replayed the failed scenarios to ensure the reproductivity of the failures. 

The scenarios selected for testing could be a threat. We counter this threat by selecting four typical and representative tasks in our evaluation. Moreover, the randomness is a threat to our results, and we mitigate the threat by repeating the experiment multiple times.
We customized SAMOTA to the Apollo+LGSVL simulation environment, which may be a threat to the results. We only customize the interface between the algorithm and the simulator without changing the algorithm. Moreover, all co-authors carefully reviewed the code. 
The diversity measurement could be another threat. To mitigate this problem, all co-authors manually study the detected critical scenarios and classify them into different categories. We uploaded videos regarding different types of violations on our website \cite{ourweb}.

\section{Related Work}
\noindent \textbf{Scenario-Based ADS Testing.}
Scenario-based testing becomes a prominent technology to evaluate the safety of ADSs, as it allows people to design complex, critical, and representative scenarios efficiently.
However, due to the complex and unpredictable environment, there are many parameters in a scenario, resulting in infinite scenario space, and it is infeasible to validate every one.
Hence, scenario-based testing focuses more on how to generate critical scenarios efficiently. 
Researchers have proposed some methods to generate critical scenarios for ADS testing, which can be categorized into data-driven methods \cite{gambi2019generating,gambi2022generating,najm2013depiction,nitsche2017pre,hauer2019did, ding2020cmts,roesener2016scenario, paardekooper2019automatic}
and searching-based methods \cite{gambi2019automatically,han2021preliminary,av_fuzzer,icse_samota,tse_adfuzz,tang2021systematic,zhou2023specification,tang2021route,tang2021collision}. 

Data-driven methods aim to generate critical scenarios from existing scenario description sources, such as accident reports and real-world motion videos.  
For example, Gambi \emph{et al.} proposed methods to generate critical simulation scenarios from police reports~\cite{gambi2019generating} and accident sketches, i.e., official crash reports with visual representations~\cite{gambi2022generating}.
Searching-based methods aim to search for critical scenarios using different technologies, such as evolutionary algorithms and guided fuzzing technologies. 
For example, the authors \cite{gambi2019automatically} applied the genetic algorithm to search for road geometries causing the failure of the AVs.
In \cite{av_fuzzer}, the authors proposed a guided fuzzer to generate critical scenarios guided by the minimal distance between the ego and NPC vehicles.
However, existing methods only consider safety-related guidance, resulting in redundant scenarios with similar motion behavior of the ego vehicle.
This paper focuses on both behavior and safety guidance to generate critical scenarios.

\noindent \textbf{ADS Specifications.}
To guarantee the safety and riding comfort of AVs, ADSs should satisfy various specifications~\cite{tuncali2019requirements, zhu2020safe}, such as safety-derived specifications (e.g.,  reachability, collision avoidance, and perception accuracy and robustness) and performance-derived ones (e.g., comfort and efficiency).
Among all specifications, safety is the most essential for ADSs~\cite{safety_first}.
Currently, scenario-based ADS testing focuses more on reachability and safety~\cite{zhong2021survey}, while perception accuracy is usually tested at model level~\cite{wang2020metamorphic}.
Hence, this paper focuses on reachability and safety specifications, mainly reaching the destination, collision avoidance, and not hitting illegal lines. 

\noindent \textbf{Scenario Evaluation.}
Assessment of the execution of a scenario is a critical step for scenario-based ADS testing. 
Some assessment metrics have been proposed in the literature~\cite{jahangirova2021quality,mahmud2017application, sharath2021literature,tuncali2019requirements}, such as time-based metrics, distance-based metrics, and deceleration-based metrics.
Recently, the quantitative semantics of signal temporal logic (STL) is applied to evaluate scenarios, which measure how far a scenario is from violating the specifications formulated in STL~\cite{dreossi2019verifai,tuncali2019requirements,tuncali2018simulation,zhou2023specification}.
Motivated by the successful application of STL in ADS testing~\cite{zhou2023specification}, in this paper, we define a violation degree to evaluate how far a scenario is from violating multiple specifications, 
which is then used as feedback to filter non-critical scenarios.

\section{Conclusion}
% Discovering diverse violations efficiently is necessary for ADS testing as the ADS usually requires a long time to run a generated scenario. 
In this paper, we propose a novel testing technique for generating diverse scenarios that cause different kinds of violations. To measure the behavior diversity of the ego vehicle, we propose \miner that collects raw states from the simulator, extracts temporal features, and performs clustering-based abstraction. To further balance the objectives of diversity and violation, we design an energy-based mechanism that is used for seed selection and mutation. Evaluation on Apollo demonstrated the effectiveness and efficiency of our method and the usefulness of \tool and the energy mechanism.
\begin{acks}
This work was partially funded by the Ministry of Education, Singapore under its Academic Research Fund Tier 1 (21-SIS-SMU-033, T1-251RES1901) and Tier 2 under Grant MOE-T2EP20120-0004.
\end{acks}

\bibliographystyle{ACM-Reference-Format}
\bibliography{bedivfuzzer}

\appendix

\end{document}